\documentclass[12pt]{iopart}

\usepackage{graphicx}
\usepackage{bm}% bold math
\usepackage{epstopdf}
\usepackage{amssymb}
\usepackage{color}
\usepackage[colorlinks=true, letterpaper=true, pdfstartview=FitV, linkcolor=blue, citecolor=blue, urlcolor=blue]{hyperref}
\usepackage{lineno}
\usepackage{longtable} 
\usepackage{booktabs}

\begin{document}

\title{The unexpected magnetism in 2D group-IV-doped GaN for spintronic applications}

\author{Rui Zhao, Rui Guo, Yiran Peng, Yanfeng Ge, Yong Liu, and Wenhui Wan*}
\address{State Key Laboratory of Metastable Materials Science and Technology $\&$ Key Laboratory for Microstructural Material Physics of Hebei Province, School of Science, Yanshan University, Qinhuangdao, 066004, P. R. China}
\ead{wwh@ysu.edu.cn}
\vspace{10pt}
\begin{indented}
\item[]December 2022
\end{indented}

\begin{abstract}
In this study, we systematically investigated the structural and magnetic properties of group-IV-doped monolayer GaN by first-principles calculations. Among the group-IV element, only Ge and Sn atoms with large atomic radii can form a buckling substituted doping structure with an in-plane magnetic moment of 1 $\mu_B$ per dopant.  The compressive strains enhance the in-plane magnetic anisotropy, while tensile strains tend to destroy the magnetic moment of dopants. Both Ge and Sn atoms can stay on the same side of monolayer GaN to form anti-ferromagnetic semiconductors due to the large diffusion barrier for crossing to the monolayer GaN. The intrinsic Ga or N vacancies will eliminate the magnetic moments of group-IV dopants due to the charge transferring from the dopants to intrinsic vacancies. The N-rich growth conditions and a plentiful supply of Ge or Sn dopants to fill the intrinsic vacancies help to maintain the magnetic properties of group-IV-doped monolayer GaN.
\end{abstract}

\section{Introduction}
It is long sought to combine magnetic and semiconducting properties within a single material for the applications of spintronics.
Transitional dilute magnetic semiconductors (DMSs) were achieved by doping group III-V and II-VI semiconductors with transition metal (TM) ions, whose partially filled $d$ or $f$ orbitals contribute to the magnetic moment \cite {b8,b9}. 
Besides that, bulk oxides of nonmagnetic (NM) cations such as ZnO and SnO$_2$ can also exhibit room-temperature ferromagnetism, namely “$d^0$ ferromagnetism”. The origin of $d^0$ magnetism was explained by native lattice
defect sites \cite{Zhang2021,Chakraborty2015} or spin-split defect impurity bands of surface defects \cite{Coey2019}.
In 2017, 2D stable magnetism were demonstrated in atomically thin CrI$_{3}$~\cite{Huang2017} and Cr$_{2}$Ge$_{2}$Te$_{6}$~\cite{Gong2017}. Based on that, 
the search for $d^0$ magnetism in 2D materials also becomes a fast-growing field and attracts much research interest. 

Until now, 
$d^0$ magnetism has been predicted in partially hydrogenated silicene \cite{Ju2016}, monolayer MoS$_2$ \cite{Gao2019}, monolayer AlN \cite{Bai2015,Han2017,a1}, and monolayer SnS$_2$ \cite{Xiao2018} doped with NM elements.   
Gallium Nitride (GaN) is a wide-band-gap semiconductor ($\sim$ 3.4 eV) with a hard and hexagonal crystal structure \cite{Adhikary2007}. GaN has been widely applied in light-emitting diodes, laser diodes, semiconductor power devices, high-frequency devices, and water-splitting devices \cite{Jabbar2021,Paula2017}. 
Compared to many TM ions which can form DMSs in doped bulk GaN \cite {b8,b9,a6}, group-IV elements exhibit as non-magnetic (NM) n-type dopants in bulk GaN \cite {a8,a4}. 
Recently, two-dimensional (2D) GaN has been synthesized with graphene as a capping sheet utilizing migration-enhanced encapsulation growth \cite{AlBalushi2016,Koratkar2016}.
2D pristine GaN is an NM semiconductor with an indirect band gap, which prevents its application in spintronics.
The doping of TM ions can induce magnetism in 2D GaN \cite {Li2018,a13,Hussain2015}. Therefore, it is straight and significant to explore the existence of the $d^0$ magnetism in 2D GaN for spintronics applications.

In the case of intrinsic defects, Ga vacancies can induce room-temperature ferromagnetism in GaN films, which has been observed in the experiment \cite {a3}. Each Ga vacancy induces a magnetic moment of 3 $\mu_B$ in monolayer GaN, which can be further tuned with the doping concentration of Mg or Si atoms \cite {a10,a12}.
N vacancies have smaller formation energy than that of Ga vacancies in both bulk and monolayer GaN \cite{Ganchenkova2006,a12}. 
There is a contradiction in the magnetic state of N vacancy. Gonz\'{a}lez et al. predicted that each N vacancy exhibited a magnetic moment of 1 $\mu_B$ \cite {a11}, while Gao et al. predicted that N vacancy was NM and could not introduce magnetism into monolayer GaN \cite{Gao2017}.
In the case of external doping, in 2015, Mu predicted that the adsorption of
F or N adatoms at low coverage made monolayer GaN become a magnetic half-metal with a high Curie temperature \cite{Mu2015}. The $2p$ orbitals of N atoms make the main contribution to the magnetic moment in F- or N-absorbed monolayer GaN \cite{Mu2015}.
In 2018, it was found that adatoms including B, C, N, Al, Si, Ga, Ge, As atoms, and O$_2$ molecular preferred to be located at top of the N atom of monolayer GaN \cite{Tang2018, Kadioglu2018}. These adatoms induce localized impurity states in the fundamental band gap of monolayer GaN, thereby bringing in the non-zero magnetic moment \cite{Tang2018, Kadioglu2018}.  
However, Kadioglu et al. further found that the aforementioned dopants were more likely to be substituted defects, as the corresponding formation energy is lower than that of adatoms' configurations \cite{Kadioglu2018}.
Unfortunately, among all the substituted defects of the aforementioned dopants, only N substitution of Ga and C substitution of N can form $d^0$ magnetism in monolayer GaN, while others are NM \cite{Kadioglu2018}. 
Here, we focus on the group-IV element including C, Si, Ge, and Sn. It was previously believed that group-IV atoms tend to substitute the Ga atoms rather than N atoms in monolayer GaN, serving as NM n-type dopants \cite {Kadioglu2018,a9,Yadav2022a}. 
We noticed that in previous works the substituted group-IV atoms were directly located at the origin position of the Ga atom in monolayer GaN \cite {Kadioglu2018,a9}. These group-IV atoms in the GaN plane can effectively hybridize with the neighboring N atoms and exhibit zero magnetic moment. However, in 2014, Gupta et al. predicted
that the substituted Si atoms in monolayer BN tend to leave the BN plane and form a bucking structure \cite{Gupta2014}, due to the larger atomic radius of the Si atom than the B atom. This kind of bucking structure weakens the hybridization between the Si atom and the neighboring N atoms, making the Si atom own an unpaired electron that produces a magnetic moment of 1 $\mu_B$ \cite{Gupta2014}.  
Since the geometry of the doping configuration is
crucial to control the magnetic properties of 2D materials, monolayer GaN with the substituted group-IV dopants in the bucking doping configuration calls for a timely investigation. The relative works have not been reported so far.

In this work, the preferential occupancy site, magnetic configuration, electronic structure, and diffusion barrier of group-IV atoms in monolayer GaN at a low doping concentration were investigated by the first-principles calculations.
We found that Ge and Sn atoms preferred to form a buckling substituted structure and exhibited a magnetic moment of 1 $\mu_B$ per dopant atom in monolayer GaN. Biaxial strain can regulate magnetic anisotropy energy (MAE) and magnetic moment of dopants. We investigated the magnetic coupling between the dopant atoms which is located at the different or same side of GaN plane. At last, the effect of intrinsic vacancies on the magnetic properties of group-IV-doped monolayer GaN was analyzed.  

\section{Computational details}
All the first-principles calculations were performed by the Vienna ab initio simulation package (VASP) \cite {b2} with the projector augmented wave (PAW) \cite {b3}  pseudopotentials and Perdew–Burke–Ernzerhof (PBE) \cite {b4} exchange-correlation functionals. 
A kinetic energy cutoff of 520 eV was adopted.
The Brillouin zone integrations were performed with $12 \times 12 \times 1$ Gamma-centered $\bf{k}$-mesh \cite{b5} for the primitive cell of
monolayer GaN.
The vacuum layer vertical to monolayer GaN was set to be 20 $\mathring{\rm A}$.
The convergence criteria of the total
energy and force were set to be $10^{-5}$ eV and 0.01 eV/$\mathring{\rm A}$, respectively.

The formation energy $E_{f}$ was defined as
\begin{equation}
E_{f}=E_{\rm GaN+X}-E_{\rm GaN}-{\mu}_{\rm Ga}n_{\rm Ga}-{\mu}_{\rm N}n_{\rm N}-{\mu}_{\rm X}n_{\rm X} \label{e1}
\end{equation}
where
$\rm E_{GaN}$ and $E_{\rm GaN+X}$ are the total energy of monolayer GaN and monolayer GaN doped with group-IV atom X (X = C, Si, Ge, and Sn), respectively. 
$n$ is the negative or positive integer that represents the number of the removed or added atoms in monolayer GaN, respectively. 
$\mu_{\rm Ga}$, $\mu_{\rm N}$, and $\mu_{\rm X}$ are the chemical potentials of the Ga, N, and group-IV atom, respectively.
We set the $\mu_{\rm X}$ as the energy of each group-IV atom in its substance. We further considered the different growth conditions.
In Ga-rich growth conditions, ${\mu}_{Ga}$ is equal to the energy of each Ga atom in the Ga substance, while ${\mu}_{n}$ is calculated by $\rm {\mu}_n={\mu}_{GaN}-{\mu}_{Ga}$ with ${\mu}_{GaN}$ be the energy per formula of monolayer GaN.
Similarly, for the N-rich growth conditions, ${\mu}_{n}$ equals the energy per N atom in N${_2}$ molecular, and ${\mu}_{Ga}$ was calculated by $\rm {\mu}_{Ga}={\mu}_{GaN}-{\mu}_{N}$.

The MAE is defined as MAE $=E_{//}-E_\perp$, where $E_{//}$ and $E_\perp$ are the energy with magnetization along the in-plane and out-of-plane   directions, respectively. 
The spin-orbit coupling (SOC) effect was considered to calculate the MAE. 
The positive and negative MAE indicates that the magnetic moment of the group IV dopant prefers to be out-of-plane and in-plane, respectively. 

\section{Results and discussion}
\subsection{Single Ge atom doping in monolayer GaN}
\begin{figure}[tbhp]
	\centering
	\includegraphics [width=0.9\textwidth]{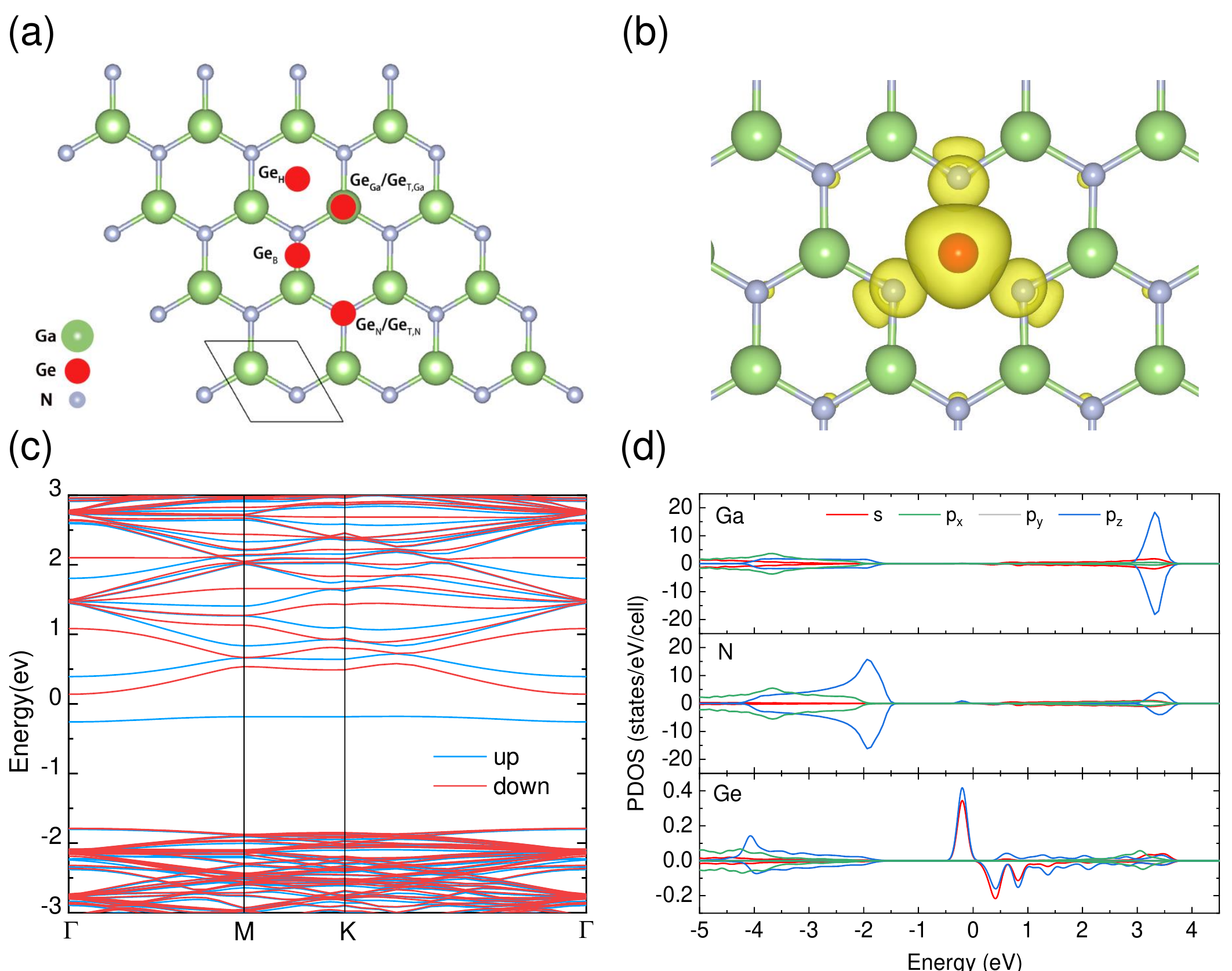}
	\caption{(a) Monolayer GaN with a Ge atom be located at different positions. (b) The spin density of ${\rm Ge}_{\rm Ga,buckling}$ structure. The isosurface is 0.002 eV/$\mathring{\rm A}^3$. The yellow color represents the positive spin density. The negative spin density is negligible. (c) Energy band and (d) projected density of states of the ${\rm Ge}_{\rm Ga,buckling}$ structure. }
    \label{fg1}
\end{figure}

Monolayer GaN adopts a planar and hexagonal lattice [see figure \ref{fg1}(a)].
We calculated the lattice constant of monolayer GaN to be 3.255 $\mathring{\rm A}$, which agrees well with previous works \cite {a3,a12,a14,a15}. The chemical bond length $\rm d_{Ga–N}$ is 1.88 $\mathring{\rm A}$. 
We first put a single Ge atom in a $6\times 6\times 1$ supercell of monolayer GaN. We considered different adsorption configurations including the Ge atom on top of the Ga site (${\rm Ge}_{\rm T, Ga}$) or the N site  (${\rm Ge}_{\rm T, N}$); Ge atom in the middle of the hexagonal ring (${\rm Ge}_{\rm H}$); Ge atom on the bridge site of Ga-N bond (${\rm Ge}_{\rm B}$) [see figure \ref{fg1}(a)]. We also examined possible substituted configurations which include the substitution of a Ga atom (${\rm Ge}_{\rm Ga, planar}$) or N atom (${\rm Ge}_{\rm N, planar}$) by a Ge atom which stays in the GaN plane; ${\rm Ge}_{\rm Ga, buckling}$ and ${\rm Ge}_{\rm N, buckling}$ represent that the substituted Ge atom was pushed out of the GaN plane by a certain bucking height.
Considering the effect of the growth conditions, we calculated the formation energy $E_{f}$ of the above doping configurations under the Ga-rich and N-rich growth conditions, respectively. Moreover, the Ge atom in the NM state and magnetic state were also considered.

\begin{table}[t]
	\caption{\label{t1}The formation energy $E_f$ (eV) and magnetic moment $m$ ($\mu_B$) of single Ge atom doped monolayer GaN in the nonmagnetic (NM) state and magnetic (M) state.}
	\begin{indented}
		\item[]\begin{tabular}{@{}lccccc}
			\br
{Position} & {Ga-rich(NM)} & {N-rich(NM)} & {Ga-rich(M)} & {N-rich(M)} &$m$ \\
			\mr
{${\rm Ge}_{\rm T, Ga}$}        & 3.82 	   & 3.82 & 3.40    & 3.40 & 2 \\
{${\rm Ge}_{\rm T, N}$}         & 2.51 	   & 2.51 & 2.31    & 2.31 & 2 \\
{${\rm Ge}_{\rm H}$}           & 3.65      & 3.65 & 3.14	& 3.14 & 2 \\
{${\rm Ge}_{\rm N, planar}$}    & 3.57 	   & 3.67 & 3.36  	& 3.46 & 1 \\
{${\rm Ge}_{\rm Ga, planar}$}   & 1.38	   & 1.28 & 1.38 	& 1.28 & 0 \\
{${\rm Ge}_{\rm N, buckling}$}  & 2.19	   & 2.29 & 2.10  	& 2.20 & 1 \\
{${\rm Ge}_{\rm Ga, buckling}$} & 1.37	   & 1.27 & 1.25  	& 1.16 & 1 \\
			\br
		\end{tabular}
	\end{indented}
\end{table}

${\rm Ge}_{\rm B}$ site is unstable at both NM and magnetic states, and will be transferred to $\rm Ge_{T, N}$ site during the structural relaxation. The results of $E_{f}$ were summarized in table \ref{t1}.
Firstly, the Ge atom in absorbed configurations can introduce a magnetic moment of 2 $\mu_B$ per Ge atom, as the corresponding $E_{f}$ of the magnetic state is smaller than that of the NM state. Meanwhile, the $E_{f}$ of absorbed configurations are not affected by growth conditions, as the $n_{\rm Ga}$ and $n_{\rm N}$ are zero in equation \ref{e1}. 
Secondly, the $E_{f}$ of substituted doping configurations 
are generally smaller than that of adsorbed configurations. 
Moreover, the Ge atom prefers to substitute the Ga atom rather than the N atom in monolayer GaN. The ground magnetic state of $\rm Ge_{Ga,planar}$ is NM, as the initial magnetic state will transition into the NM state.
The above results agree well with previous works  \cite{Kadioglu2018,a9,Yadav2022a}, indicating the reliability of our calculations. 
Thirdly, compared to the planar ${\rm Ge}_{\rm Ga,planar}$ configuration, Ge atom in the buckling substituted doping configuration ${\rm Ge}_{\rm Ga,buckling}$ has a lower $E_f$ in both NM and magnetic states, thereby being the most stable doping configuration.
This can be directly reflected by the evolution of energy of ${\rm Ge}_{\rm Ga}$ as the Ge atom gets close to the Ga vacancy gradually, as shown in figure \ref{a1}(a).  
Three valence electrons of the Ge atom in ${\rm Ge}_{\rm Ga,buckling}$ structure participate in the chemical bonding with three neighboring N atoms, leaving a localized and unpaired electron on the Ge atom which contributes a magnetic moment of 1 $\mu_B$. The $E_f$ of ${\rm Ge}_{\rm Ga,buckling}$ can be further reduced in the N-rich growth conditions, as shown in table \ref{t1}. 
Thus, Ge atoms indeed can induce non-zero magnetic moment and magnetic properties in monolayer GaN.  
However, the doping of the Ge atom into monolayer GaN is an endothermal reaction that needs extra energy to promote the doping process. 

We examined the lattice structure and electronic structures of ${\rm Ge}_{\rm Ga,buckling}$ configuration. The bucking height $h$ of Ge dopant is about 0.6 \AA.
The optimized Ge-N bond length is 0.01 \AA\ larger than that of Ga-N bonding. Because the neighboring N atoms get close to the Ge dopant along the in-plane direction, the neighboring Ga-N bond lengths were enlarged by 0.01 \AA.
The spin density of ${\rm Ge}_{\rm Ga,buckling}$ is mainly distributed around the Ge atom and its three nearest neighboring N atoms  [see figure \ref{fg1}(b)]. 
Compared to the positive spin density, the negative spin density is negligible.  
Figure \ref{fg1}(c) displays the band structure of ${\rm Ge}_{\rm Ga,buckling}$ configuration which still keeps the semiconducting properties. The top valence band and bottom conduction band are comprised of spin-down and spin-up electronic states, respectively. 
Ge dopant introduces a dispersionless impurity band among the band gap 
 and other impurity bands near the conduction bands of monolayer GaN.
Thus, ${\rm Ge}_{\rm Ga,buckling}$ configuration has a band gap of about 0.40 eV at the PBE level [see figure \ref{fg1}(c)].
The $s$, $p_z$ orbital of the Ge atom and the $p_z$ orbital of the neighboring N atoms around the Ge atom dominate the electronic states around the Fermi level, as shown in the projected density of states (PDOS) in figure \ref{fg1}(d). 
Through the analysis of the orbital-resloved magnetic moment, we found that the same orbitals dominate the magnetic moment of ${\rm Ge}_{\rm Ga,buckling}$ configuration. On the other side, the contribution of Ga atoms to the mganetic moment can be negligible. 

We further compared the energy of ${\rm Ge}_{\rm Ga,buckling}$ configuration with magnetic moment along different directions including
[100] [010] [001] [110] [111] directions [see table \ref{t3.32}].
The magnetic moment of Ge dopant prefer the in-plane direction. 
The negative MAE is estimated to be about -14.55 $\mu$eV per Ge atom. 
Moreover, the energy difference of MAE is less than 0.1 $\mu$eV for magnetic moment lying in the $xy-$plane.
Thus, the MAE of ${\rm Ge}_{\rm Ga,buckling}$ can be seen to be isotropic in the $xy-$plane.  

\begin{figure}[tbhp]
	\centering
	\includegraphics[width=0.9\textwidth]{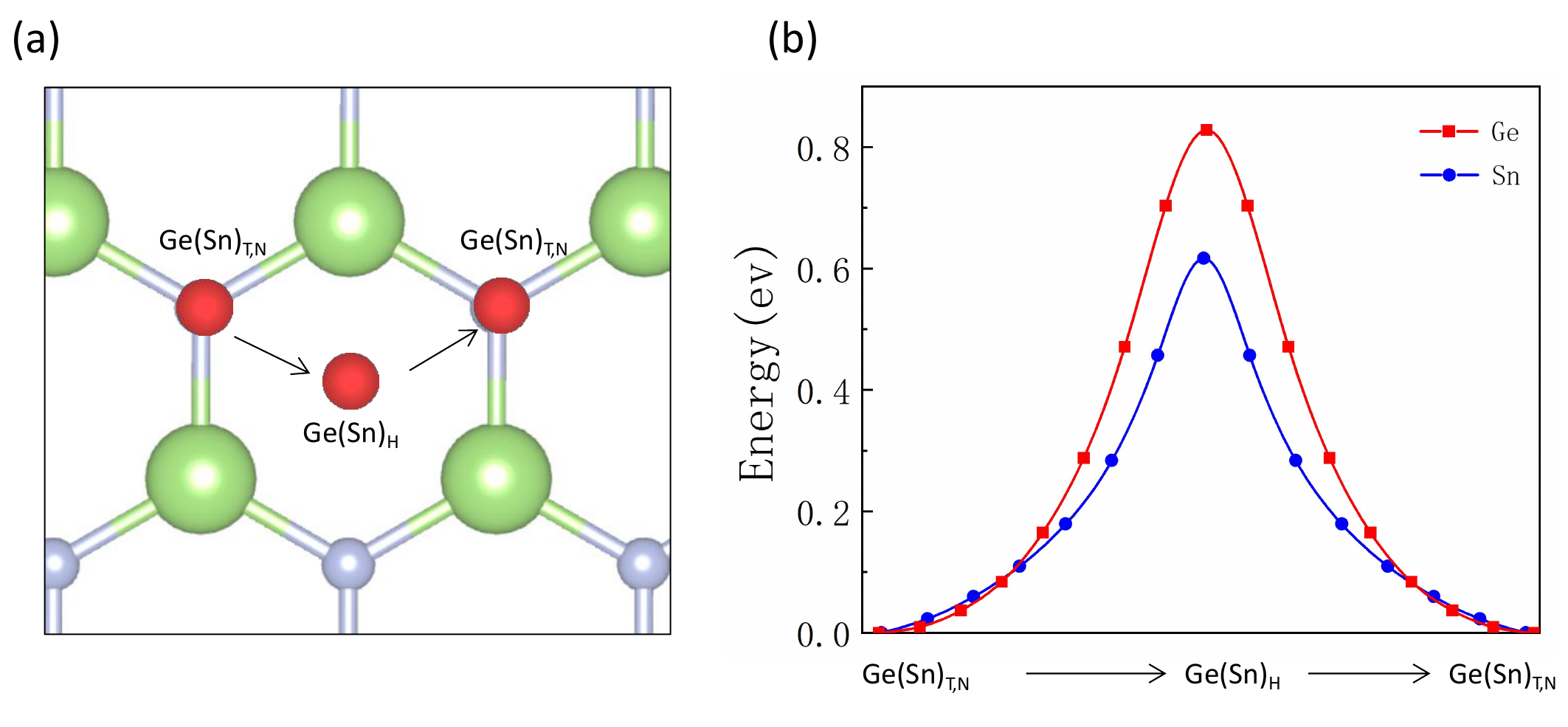}
	\caption{(a) Diffusion path and (b) corresponding energy profile of single Ge and Sn atoms along the surface of monolayer GaN.}
   \label{fg2}
\end{figure}

We also investigated the diffusion of the Ge atom crossing or along the GaN plane. In the vertical direction, there are two symmetry equivalent ${\rm Ge}_{\rm Ga,buckling}$ configuration on both sides of monolayer GaN. The Ge atom can cross the GaN plane and reach the other ${\rm Ge}_{\rm Ga,buckling}$ site. 
By the nudged elastic band (NEB) method \cite{Mills1995}, the vertical diffusion barrier $\bigtriangleup E$ was identified to be the energy difference between the ${\rm Ge}_{\rm Ga,buckling}$ and ${\rm Ge}_{\rm Ga,planar}$, which is about 0.12 eV. 
On the other side, the diffusion of the Ge atom parallel to the GaN plane can proceed through the low-energy adsorption sites,
which is ${\rm Ge}_{\rm T, N}-{\rm Ge}_{\rm H}-{\rm Ge}_{\rm T, N}$ pathway [see figure \ref{fg2}(a)]. Through the
NEB method \cite{Mills1995}, the optimized saddle point was near the ${\rm Ge}_{\rm H}$ site. The in-plane diffusion barrier $\bigtriangleup E$ was estimated at 0.82 eV [see figure \ref{fg2}(b)].

\subsection{Single group-IV atom doping in monolayer GaN}
We expanded current calculations to the doping of other group-IV atom X (X = C, Si, Sn) in monolayer GaN. 
Both ${\rm C}_{\rm Ga,buckling}$ structures and ${\rm Si}_{\rm Ga,buckling}$ structures are not stable and will transition into the NM planar doping configuration after structural relaxation.
The C and Si atoms have smaller atomic radii than that of the Ga atom. Their valence electrons can effectively participate in the bonding with neighboring N atoms when they stay in the GaN plane. Meanwhile, all the valence electrons become nonlocalized bonding electrons, so no magnetic moment appears. Their buckling doping structure will weaken the C-N or Si-N bonding, so has higher energy than that of planar doping configuration. 
As a result, both of them can not introduce magnetism in monolayer GaN.  

Similar to the Ge dopant, Sn atom prefers the bucking doping structure ${\rm Sn}_{\rm Ga,buckling}$ with a magnetic moment of 1 $\mu_B$ per Sn atom [see figure \ref{a1}(b)]. 
The corresponding $E_f$ is lower in the N-rich growth conditions [see table \ref{t2}]. Thus, the Sn atom is also suited for inducing magnetic properties into monolayer GaN. 
The Ge and Sn atoms have large atomic radii and feel the strain in the ${\rm X}_{\rm Ga, planar}$ structure. Their buckling doping structures can effectively release the strain, thereby decreasing the total energy. The buckling structure offers a $sp^3$-like bonding environment. A valence electron of Ge or Sn becomes an unpaired and localized electron which produces a magnetic moment of 1 $\mu_B$. 

\begin{table}[t]
	\caption{\label{t2}The formation energy $E_f$ (eV) of single Sn atom doped monolayer GaN in nonmagnetic (NM) state and magnetic (M) state.}
		\begin{indented}
			\item[]\begin{tabular}{@{}lcccc}
				\br
				{Position} & {Ga-rich(NM)} & {N-rich(NM)} & {Ga-rich(M)} & {N-rich(M)} \\
				\mr
				{${\rm Sn}_{\rm T,Ga}$}           & 3.30  &3.30 &2.91 & 2.91  \\
				{${\rm Sn}_{\rm T,N}$}            & 2.58  &2.58 &2.06 &2.06  \\
				{${\rm Sn}_{\rm H}$}              & 3.15  &3.15 &2.68 &2.68  \\
				{${\rm Sn}_{\rm N,planar}$}       &4.90   &5.00 &4.71 & 4.82 \\
				{${\rm Sn}_{\rm Ga,planar}$}      &2.16   &2.06 &2.16 &2.06   \\
				{${\rm Sn}_{\rm N,buckling}$}     &2.23   &2.33 &2.11 &2.21  \\
				{${\rm Sn}_{\rm Ga,buckling}$}    &1.77   &1.67 &1.62 &1.52  \\
				\br
			\end{tabular}
		\end{indented}
	\end{table}

The bucking height of the Sn atom in ${\rm Sn}_{\rm Ga,buckling}$ is about 0.9 $\mathring{\rm A}$, larger than that of the Ge atom. 
Similarly, the spin density is mainly located on the Sn atom and its three nearest N atoms. The Sn dopant in ${\rm Sn}_{\rm Ga,buckling}$ structure introduces two dispersionless impurity bands with different spins in the fundamental band gap of monolayer GaN [see figure \ref {a2}(a)].
The splitting of the spin-up and spin-down bands causes ${\rm Sn}_{\rm Ga,buckling}$ to be a semiconductor with a band gap of 0.58 eV.
The PDOS in figure \ref {a2}(b) shows that the magnetic moment is dominated by the $s$ and $p_{z}$ orbitals of the Sn atom as well as the $p_{z}$ orbital of the neighboring N atoms.
As shown in table \ref{t3.32}, the magnetic moment of Sn dopant also prefers to be in-plane. 
The MAE is almost isotropic and is about -121.27 $\mu$eV per Sn atom, which is larger than that of Ge dopant due the stronger SOC of Sn atom than Ge atom. 

The potential barrier $\bigtriangleup E$ to be overcome by the Sn atom crossing the GaN plane was estimated to be about 0.54 eV by the NEB method \cite{Mills1995}, which is larger than that of Ge atoms. The Sn atoms are less likely to cross the GaN plane than the Ge atoms.
The in-plane diffusion of the Sn atom proceeds with the same diffusion path as that of the Ge atom. The corresponding $\bigtriangleup E$ is estimated as about 0.62 eV [see figure \ref{fg2}(b)], smaller than that of the Ge atom.

\begin{table}[htb!]
	\centering
	\caption{\label{t3.32}The MAE ($\mu$eV per dopant atom) with magnetization along the different directions.}
	\begin{tabular}{lcccccccc}
		\br
		& [100] & [010]& [001]& [110]&[111] \\
		\mr
		Ge & 	   0&     0 &      14.55 & -0.02& 4.92   \\
		Sn& 	  0 & -0.05 &     121.27 & -0.08& 40.99  \\
		\br
	\end{tabular}
\end{table}

\subsection{Strain effect}

\begin{figure}
	\centering
	\includegraphics[width=0.8\textwidth]{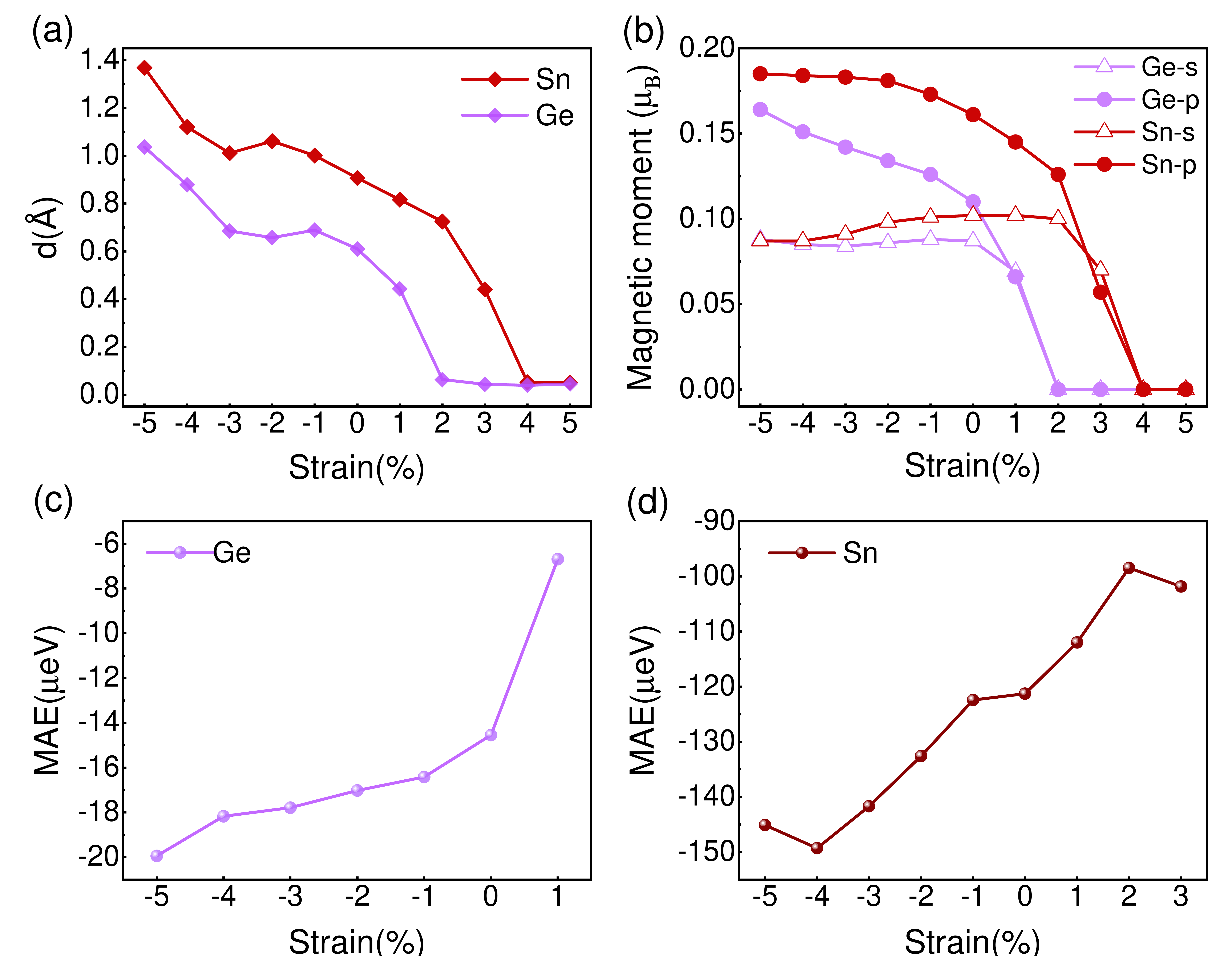}
	\caption{(a) The buckling height $d$ and (b) orbital resolved magnetic moment as a function of strain. (c) and (d) are the MAE of ${\rm Ge}_{\rm Ga,buckling}$ and  ${\rm Sn}_{\rm Ga,buckling}$ at different strains, respectively.
	}
	\label{fg3}
\end{figure}

Strain is a effective way to modulate the structural and magnetic properties of 2D materials \cite{Fan2022,wang2023,shengmei2021,yan2021,zhang2020}.
We defined the strain as $\varepsilon=\frac{(a-a_0)} {a_{0}} \times 100\%$ , where $a$ and $a_{0}$ are the lattice constants of the strained and unstrained structures, respectively.
Compressive strain decreases the lattice constants of the ${\rm Ge}_{\rm Ga,buckling}$ and ${\rm Sn}_{\rm Ga,buckling}$ structures, but increases the buckling height $d$ [see figure \ref{fg3}(a)]. The tensile strain has the opposite effect. 
The stretching of the lattice increases the space of the Ga vacancy, so that the Ge or Sn dopant gradually falls into the GaN plane.
The ${\rm Ge}_{\rm Ga,buckling}$ and ${\rm Sn}_{\rm Ga,buckling}$ structures transfer to in-plane substituted structures at a tensile strain of 2\% and 4\%, respectively. Meanwhile, the magnetic moments of Ge and Sn dopant disappear at the corresponding tensile strains [see figure \ref{fg3}(b)].
As long as $d$ does not approach 0, the substituted doping structure always has magnetic moment, indicating a potential application in strain-switched spintronic devices.   

\begin{figure}
	\centering
	\includegraphics[width=0.9\textwidth]{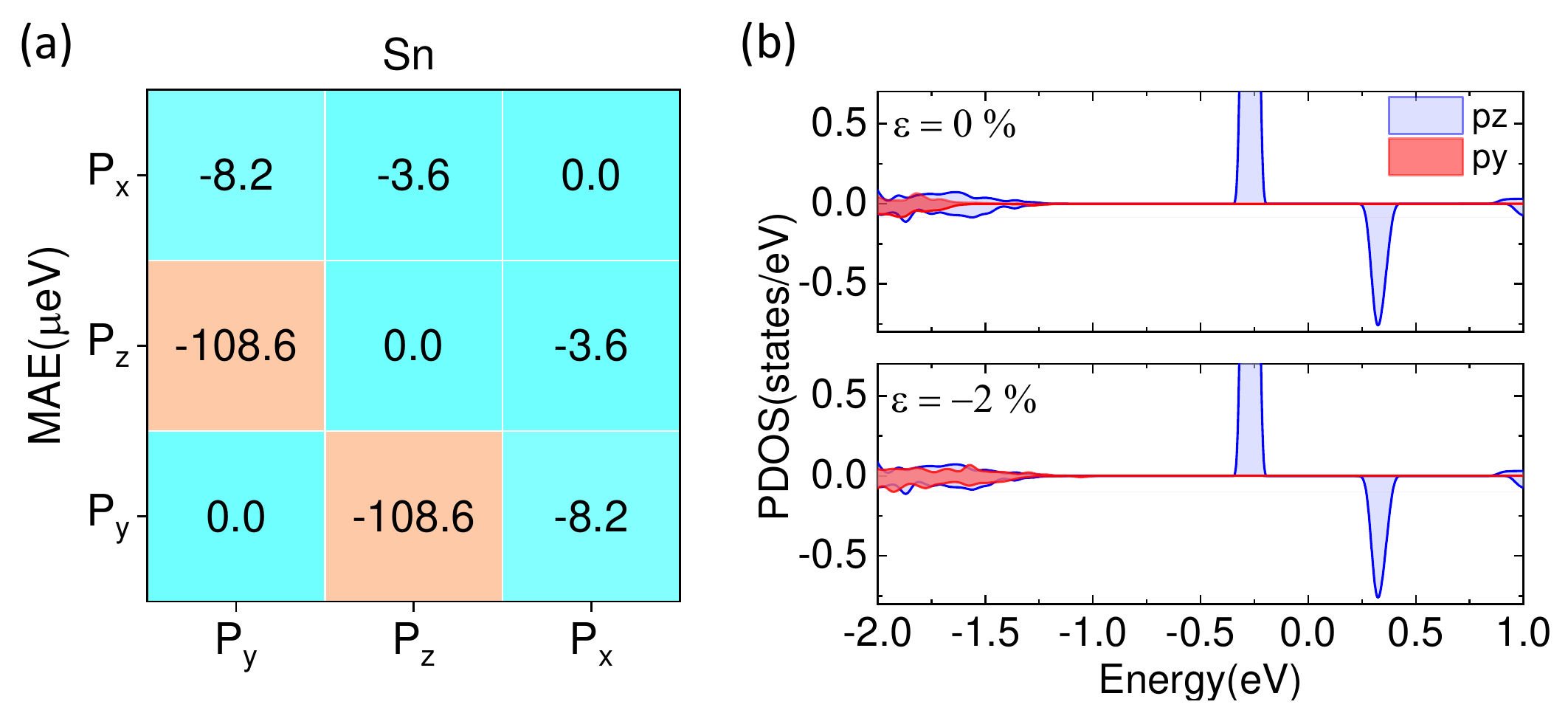}
	\caption{(a) orbital-resolved MAE of ${\rm Sn}_{\rm Ga,buckling}$ at strain of $\varepsilon=0\%$. (b) The contribution of Sn-$p_y$ and $p_z$ orbital to the PDOS of ${\rm Sn}_{\rm Ga,buckling}$ $p$ orbitals at a strain of 0\% and -2\%.}
	\label{fg4}
\end{figure}

The effect of strain on the MAE of Ge and Sn dopant are shown in figure \ref{fg3}(c) and \ref{fg3}(d). MAE keeps to be negative at strain from -5\% and 5\%. 
Compressive and tensile strain will decrease and increase the MAE, respectively. Thus, compressive strain help to enhance the in-plane magnetic anisotropy of ${\rm Ge}_{\rm Ga,buckling}$ and ${\rm Sn}_{\rm Ga,buckling}$ structures. 
To understand the physics mechanism behind the strain modulation of MAE, we displayed the orbital-resolved MAE of ${\rm Sn}_{\rm Ga,buckling}$ in figure \ref{fg4}(a). 
Based on the second-order perturbation theory \cite{PRB1993}, the MAE can be represented by 
\begin{equation}
	{\rm MAE} = 
	 \zeta^2  \sum_{o,\alpha,u,\beta} (2\delta_{\alpha\beta}-1) \frac{|\left \langle {{o^\alpha}} \right|{\hat L_z}\left| {u^\beta} \right\rangle |^2 -
		|\left\langle {{o^\alpha}} \right|{\hat L_x}\left| {u^\beta} \right\rangle |^2}{E^{\alpha}_{u}-E^{\beta}_{o}} .  
	\label{e3.32}
\end{equation}
Here $\zeta$ is the SOC strength. The $\alpha$, $\beta$ represent spin-up ($\uparrow$) or spin-down ($\downarrow$) state.
$E_o$ and $E_u$ represent the energy of occupied ($o$) and unoccupied ($u$) states, respectively.
The MAE was determined by the SOC matrix elements, the energy difference between the $o$ and $u$ states, and the density of $o$ ($u$) states.
The matrix element in the numerator of Eq.~\ref{e3.32} is nonzero for
$\left\langle {m} \right|{\hat L_z}\left| {m} \right\rangle $ and
$\left\langle {{m}} \right|{\hat L_x}\left| {m\pm1} \right\rangle $, where $m$ is the magnetic quantum number.
Thus, the matrix element between $p_z$ orbital is zero, though Sn-$p_z$ orbital dominates the PDOS near the Fermi level [see figure \ref{a2}(a)].
Combined the figure \ref{fg4}(a) and \ref{fg4}(b), the matrix element $\left\langle {p^{\downarrow}_{y}} \right|{\hat L_x}\left| p^{\downarrow}_{z} \right\rangle $ between the Sn-$p_y$ state in the valence bands and the Sn-$p_z$ state in the conduction bands make the main contribution to the negative MAE of ${\rm Sn}_{\rm Ga,buckling}$. 
Compressive strain enhances the $\sigma$-type bonding between the $p_{x}$ and $p_{y}$ orbital of Sn and neighboring N atoms.
Since the energy of Sn-$p$ orbital is higher than that of N-$p$ orbital, 
the enhanced hybridization will further increase the energy of $p_{x}$ and $p_{y}$ orbital of Sn atom in the valance bands. 
As a result, the energy difference between the Sn-$p_y$ state and Sn-$p_z$ state decrease under compressive strain, as shown in see figure \ref{fg4}(b).
The decrease of denominator ${E_{u}-E_{o}}$ of equation \ref{e3.32} leads to the decrease of the MAE of ${\rm Sn}_{\rm Ga,buckling}$ under compressive strain [see figure \ref{fg3}(d)]. 
The strain modulation of MAE of ${\rm Ge}_{\rm Ga,buckling}$ can be explained by similar analysis.
Generally, ${\rm Sn}_{\rm Ga,buckling}$ has a larger SOC effect and a larger absolute value of MAE than that of ${\rm Ge}_{\rm Ga,buckling}$.
Meanwhile, the magnetic moment of Sn dopant can sustain under a larger range of external strain than the Ge dopant [see figure \ref{fg4}(a)].

\subsection{Magnetic coupling between group-IV dopant in monolayer GaN}
\begin{figure}
	\centering
	\includegraphics[width=0.9\textwidth]{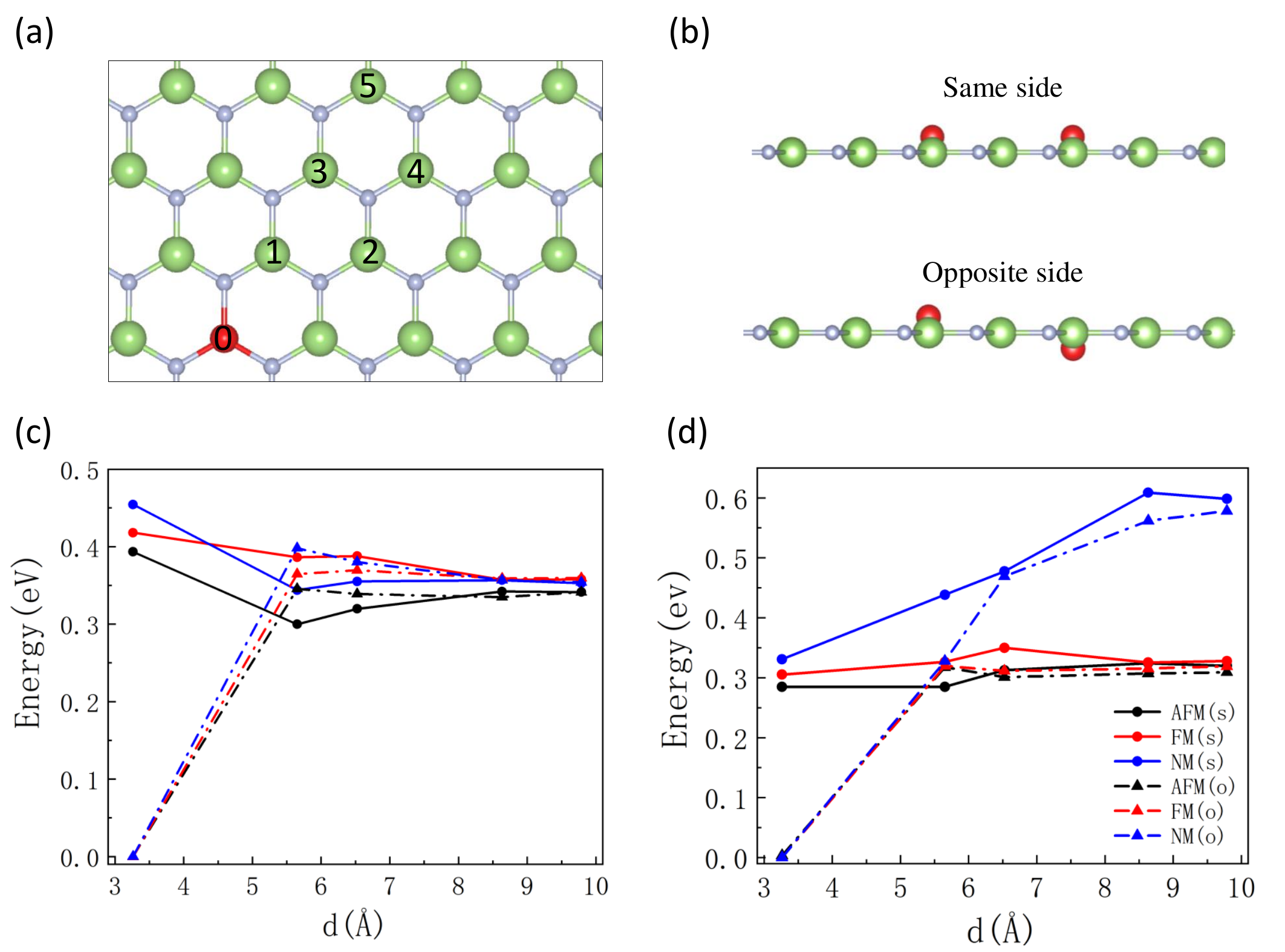}
	\caption{(a) The doping position of double Ge atoms in monolayer GaN. One is located at site 0, the other one is located at $1\sim 5$ site. (b) The Schematic diagram of two doped Ge atom on the same (S) and opposite (O) side of monolayer GaN. The total energy as a function of distance between the doped (c) Ge atom and (d) Sn atoms. The energy of doped monolayer GaN with the nearest dopant distance was set to 0.}
	\label{fg5}
\end{figure}

To study the magnetic coupling between the group-IV dopants at high doping concentration, we substituted two Ga atoms with Ge or Sn atoms at different distances in a 10$\times$10$\times$1 supercell of monolayer GaN [see figure \ref{fg5}(a)]. 
Meanwhile, we considered two doped atoms on the same or different sides of monolayer GaN [see figure \ref{fg5}(b)].
The possible magnetic coupling between two dopants includes ferromagnetic (FM), NM, and antiferromagnetic (AFM) coupling. 

As shown in the figure \ref{fg5}(c, d), the doping configuration with the lowest energy is that two doped Ge or Sn atoms are located at opposite sides of monolayer GaN with the nearest distance. The corresponding doped configuration was labeled as (0, 1) which represent that a Ge atom is at the 0 site and the other one is located at the 1 site, as shown in figure \ref{fg5}(a). Meanwhile, we found that both FM and AFM states of the configuration (0, 1) were converted to the NM state. 
Therefore, when the distance between the group-IV dopants is too small, the aggregation phenomenon will occur and the magnetic moments of group-IV atoms disappear. However, we have found that the  Ge and Sn atom have a large diffusion barrier $\bigtriangleup E$ vertical to the GaN plane. Especially, the $\bigtriangleup E$ of the Sn atom reaches 0.54 eV. Thus, if we grew monolayer GaN on a suitable substrate and doped it with Ge or Sn atoms from  one side of monolayer GaN, most of dopants might stay on the same side of monolayer GaN. 
For Ge or Sn atoms on the same side of monolayer GaN, the (0, 2) configuration in the AFM state has the lowest energy, as shown in figure \ref{fg5}(c, d). The dopant-dopant distance is about 5.7 \AA.
As the distance between two doped atoms increases, the magnetic coupling between them gradually decreases. The difference between the energy of FM and AFM state becomes small at a large distance.

\begin{figure}
	\centering
	\includegraphics[width=0.9\textwidth]{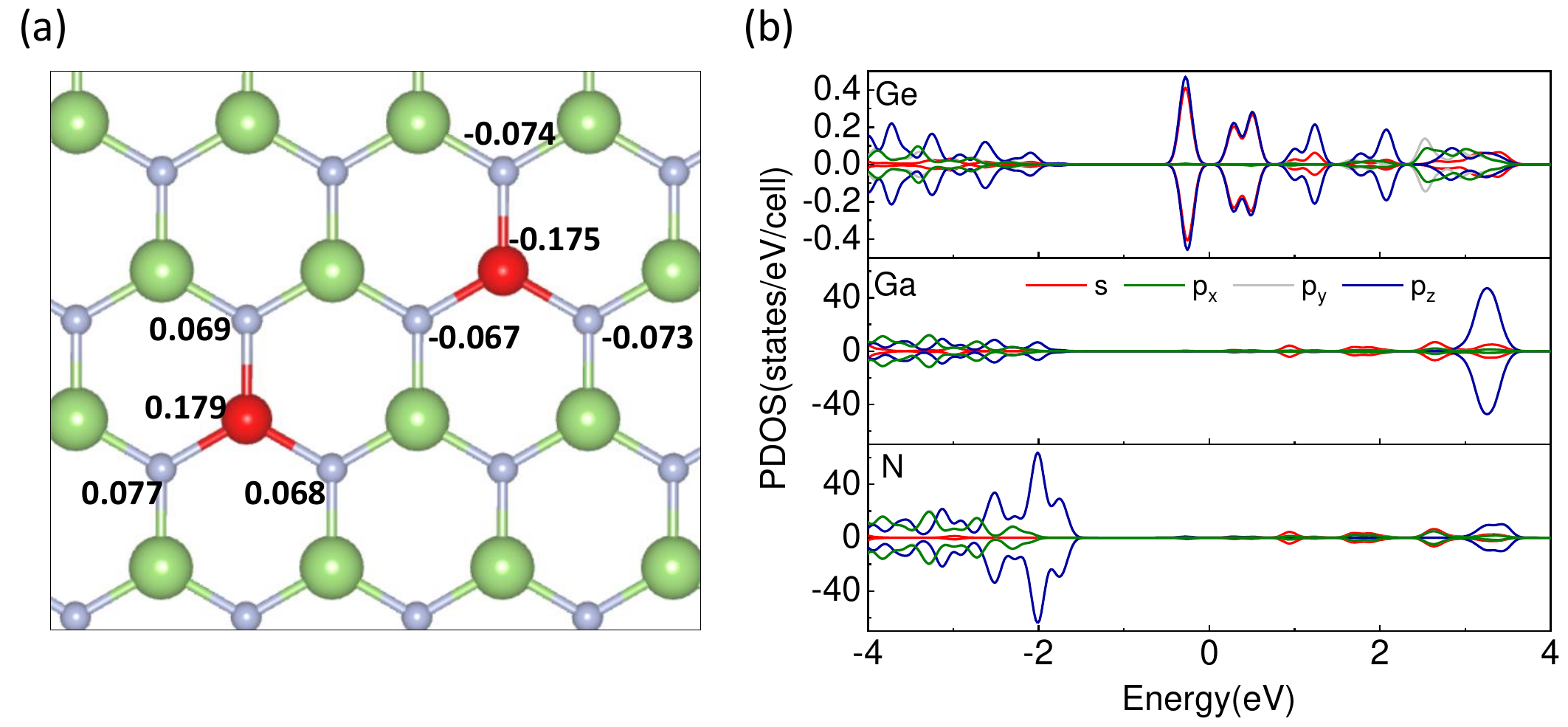}
	\caption{(a) The distribution of magnetic moment of monolayer GaN doped with double Ge atoms in the antiferromagnetic state.  (b) The corresponding PDOS of doped configuration in (a).}
	\label{fg6}
\end{figure}

Figure \ref{fg6}(a) displays the spin distribution of the configuration (0, 2) in the AFM state. 
It can be seen that the doped Ge atoms and their nearest N atoms provide the largest contribution to the magnetic moments. The direction of magnetic moments is the same for Ge and its neighboring N atoms, but is different for two doped Ge atoms. The corresponding PDOS in figure \ref{fg6}(b) shows a symmetric DOS in the spin-up and spin-down channels. 
Through the checking of the PDOS of monolayer GaN with two doped Ge atoms on the same side, we found that all the double doped GaN exhibited AFM semiconducting propeties [see figure \ref{fg6}(b)]. The $s$ and $p_z$ orbitals of the Ge atoms dominate the PDOS at the Fermi level. The representative spin distribution and PDOS of the AFM state of monolayer GaN with two doped Sn atoms is shown in figure \ref{a4}, which also exhibits similar AFM semiconducting properties. 

We noticed that if one fully substituted Ga atoms by Ge or Sn atoms, one can obtained monolayer GeN and SnN which are FM semiconductors with large band gap and high Curie temperatures \cite{a1,Tkachenko2020}. Thus, there should be an AFM-to-FM magnetic transition in group-IV doped monolayer GaN as the doping concentration increases, which can be explained it by the enhanced exchange splitting and delocalized impurity states of group-IV dopants \cite{a1}. 

\subsection{The influence of intrinsic vacancy on the group-IV-doped monolayer GaN}
At last, we considered the effect of intrinsic vacancies on the magnetic properties of group-IV-doped monolayer GaN. We first created a Ga vacancy ($\rm V_{Ga}$) or N vacancy ($\rm V_{N}$) in a $6\times 6\times 1$ GaN supercell and calculated its magnetic state. We found that the size of the $\bf k$-grid mesh was crucial to appropriately predict the ground magnetic state of intrinsic vacancies of monolayer GaN. For example, one might predict $\rm V_{N}$ to be a NM state with a course $\bf k$-grid mesh. However, $\rm V_{N}$ should be in the magnetic state with a magnetic moment of 1 $\mu_B$ using a dense $\bf k$-grid mesh. That can explain the contradiction in previous works \cite{Gao2017,a11}. 
Next, we substituted one Ga atom in the $6\times 6\times 1$ GaN supercell, 
and two Ga atoms in a $10\times 10\times 1$ supercell of by group-IV atom X (X = Ge, Sn). The group-IV dopants are in the stable ${\rm Ge}_{\rm Ga,buckling}$ or ${\rm Sn}_{\rm Ga,buckling}$ configuration. 
For monolayer GaN with two doped Ge or Sn atom, we choose the configuration (0, 2) as the representative configuration [see figure \ref{fg5}(a)].
Meanwhile, two Ge or Sn atoms are in the AFM configuration and locate on the same side of monolayer GaN.
Then we created a Ga or N vacancy ($\rm V_{Ga}$) at different distance $d$ between the vacancy and the dopants in the supercell of monolayer GaN.  

\subsubsection{N vacancy}
For the case of N vacancy, we found that group-IV-doped monolayer GaN containing a N vacancy always exhibit the NM state. Thus, N vancancy is not desirable for magnetic propreties of group-IV-doped monolayer GaN. 
According to previous content, the N-rich growth conditions are more favorable for not only substituting of the Ga atoms by group-IV atoms but also preventing the forming of N vacancies in monolayer GaN. Cui et al. had proposed that the disproportionation of NO can be used to repair the N-vacancy of monolayer InN \cite{Cui2019}. Similar methods can be applied in monolayer GaN to decrease the concentration of N vacancies in the future.  

\subsubsection{Ga vacancy}
\begin{figure}
	\centering
	\includegraphics[width=0.9\textwidth]{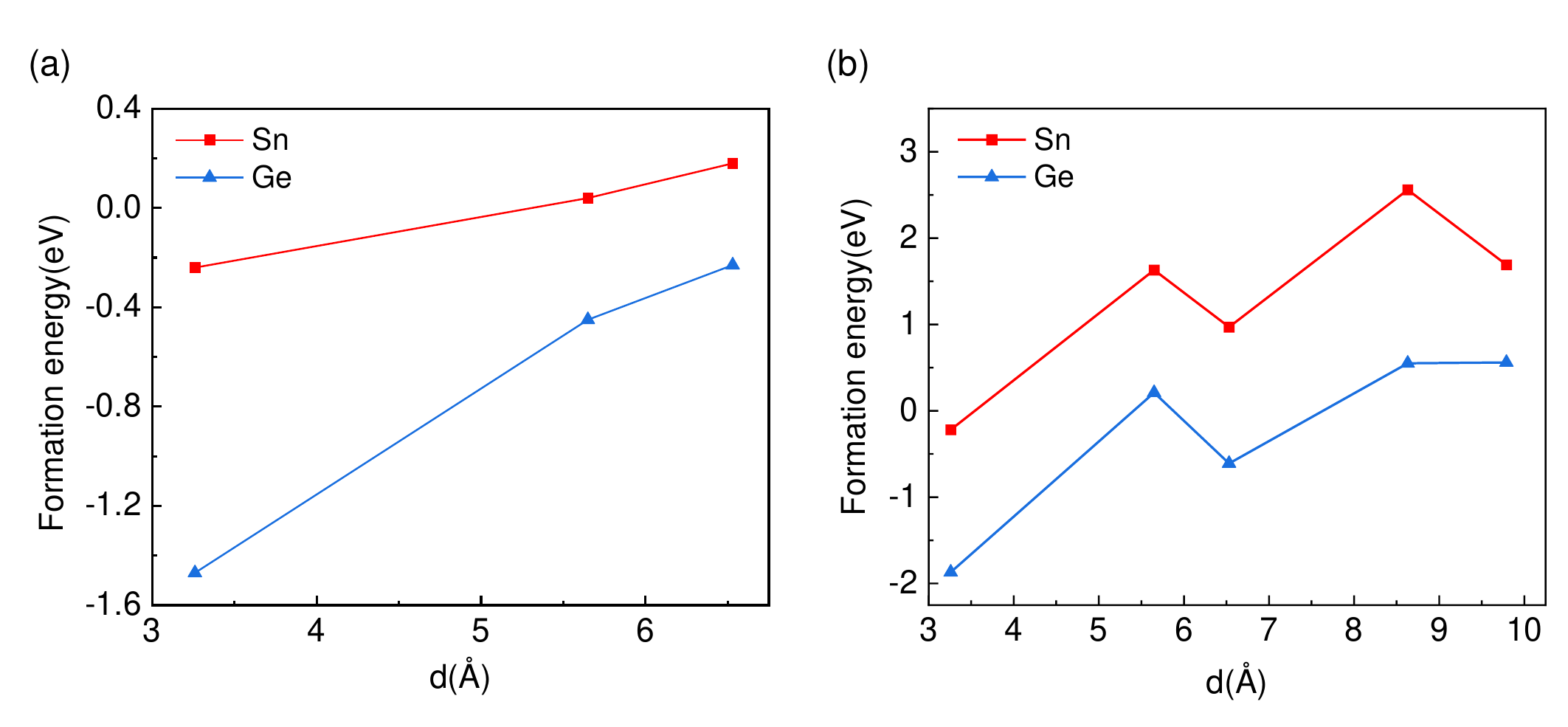}
	\caption{The formation energy of group-IV-doped monolayer GaN with 
		a Ga vacancy (a) $\rm V_{Ga} + X$ and $\rm V_{Ga} + 2X$ as a function of distance between the dopant and Ga vacancy in the N-rich growth  conditions. X present the group-VI dopant Ge and Sn.}
	{\label{fg7}}
\end{figure}

We labeled group-IV-doped monolayer GaN with a Ga vacancy and one or two dopants as $\rm V_{Ga} + X$ or $\rm V_{Ga} + 2X$ configurations, respectively.
As shown in figure \ref{fg7} (a, b), the relevant $E_f$ of $\rm V_{Ga}+X$ or $\rm V_{Ga}+2X$ are even negative for small $d$. The $E_f$ in general increases as with the $d$ increasing. That indicated that the $\rm V_{Ga}$ tends to get close to Ge or Sn dopants to form complex structures.

\begin{figure}
	\centering
	\includegraphics[width=0.9\textwidth]{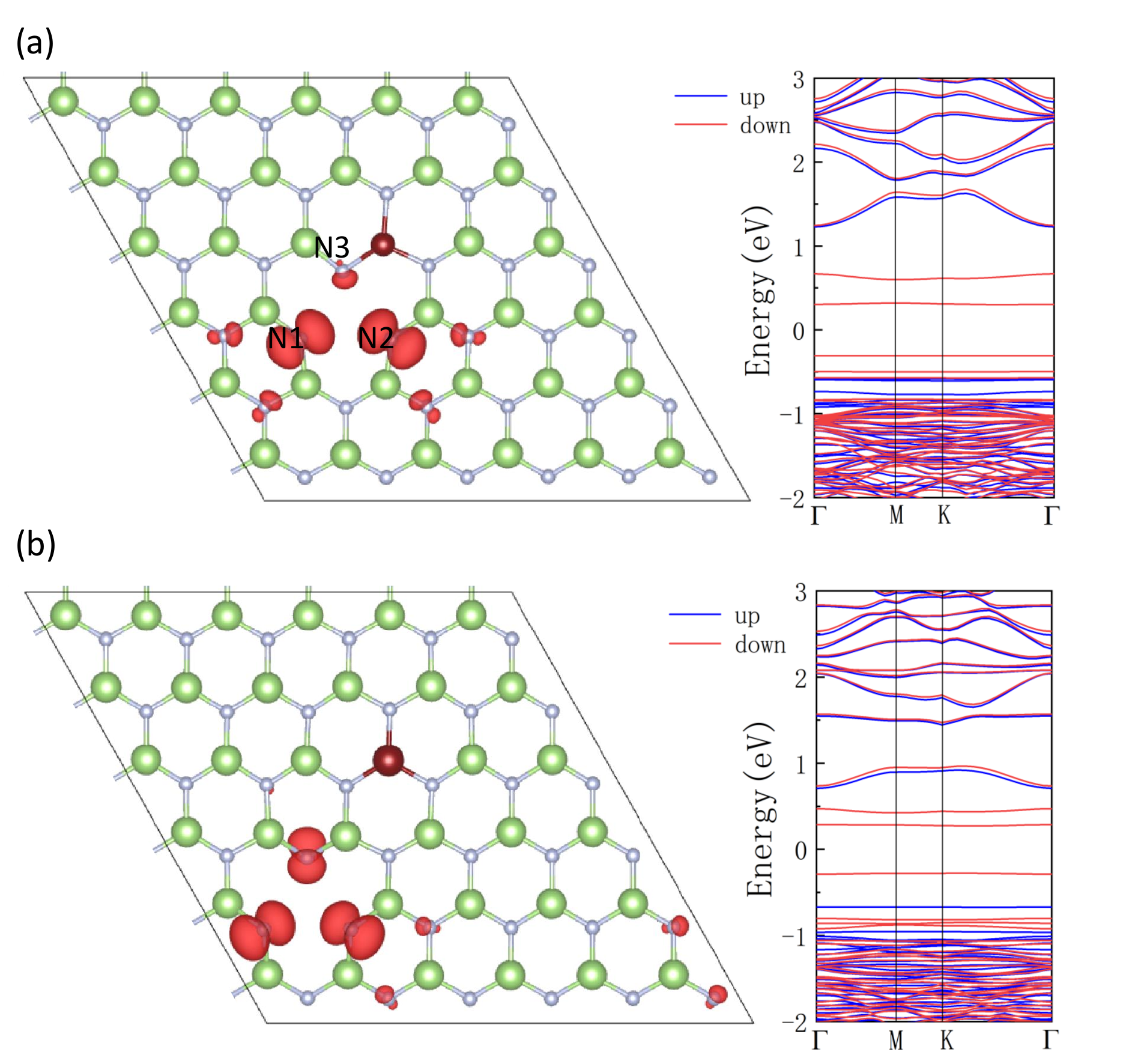}
	\caption{Spin density and band structure of $\rm V_{Ga}-Sn$ with (a) small and (b) large distance between dopant and vacancy. Positive and negative spin density are labeled as red and yellow, respectively. The spin-equivalent surface is 0.0035 eV/$\mathring{\rm A}^3$.}
	{\label{fg8}}
\end{figure}

The representative distribution of spin density and band structures 
for $\rm V_{Ga}+Sn$ at different $d$ are displayed in figure \ref{fg8} (a, b). 
Moreover, the band gap is insensitive to $d$.
A similar situation was also founded in $\rm V_{Ga}+Ge$ structures.
The band gap of $\rm V_{Ga}+Ge$ or $\rm V_{Ga}+Sn$ at different $d$ is about 1.00 eV and 0.60 eV, respectively. 
Though $\rm V_{Ga}+Sn$ also exhibit the spin-polarized band structure,   
the spin density is mainly distributed on the N atoms near the Ga vacancy. 
The spin density around the group-IV dopant is negligible. 
The similar situation happens in $\rm V_{Ga} + 2X$ configurations.
Through the examination of the lattice structure of $\rm V_{Ga} + X$ or $\rm V_{Ga} + 2X$ configurations, we found that the group-IV dopant returned to the GaN plane in the presence of V vacancy. That means that all the valence electrons of group-IV dopants become bonding electrons and can not exhibit magnetic moments.

Each Ga vacancy in monolayer GaN leaves three unpaired electrons on the three neighboring N atoms around the Ga vacancy, leading to the magnetic moment of 3 $\mu_B$ \cite{Gao2017}. When another Ga atom was replaced by Ge or Sn atom, an extra electron was brought in monolayer GaN. However, at this moment, we found that the extra electron could not keep around the dopant as the unpaired electron, but transfer to the N atoms near the Ga vacancy to fill a hole. As a result, the magnetic moment of $\rm V_{Ga}+Ge$ or $\rm V_{Ga}+Sn$ is only 2 $\mu_B$. 
The transferring of charge was found even in the $\rm V_{Ga} + X$ configurations with a large $d$. 
Similarly, we found that the magnetic moments of most $\rm V_{Ga}+2X$ configurations are 1 $\mu_B$, due to that two extra electrons transfer from two group-VI dopants to the N atoms near the Ga vacancy.
At the moment, Ge or Sn atoms were only used to tune the magnetic moment of Ga vacancy by changing the hole number, rather than induce magnetic moment.
Therefore, the Ga vacancy will also remove the unpaired electron on the Ge or Sn dopant, which is not desirable for the magnetic properties of group-IV doped monolayer GaN. 
 
How to avoid such situation? Beginning with monolayer GaN with a Ga vacancy ($\rm V_{Ga}$), we compared the formation energy $E_f$ of two process. One is that Ge or Sn atom directly fulfills the Ga vacancy, and the other one is that Ge or Sn atom substitutes another Ga atom in the monolayer GaN. 
The $E_f$ of former process is much smaller than the later one. 
Thus, the doping group-IV atoms are more inclined to fill the intrinsic Ga vacancies directly. 
If one supplied enough concentration of Ge or Sn dopants into monolayer GaN in the N-rich growth conditions, 
most Ge or Sn atoms will mainly form the ${\rm Ge}_{\rm Ga,buckling}$ or ${\rm Sn}_{\rm Ga,buckling}$ configurations.
Meanwhile, the forming of intrinsic Ga and N vacancies can be effectively suppressed, which helps to realize the introduction of magnetic properties into monolayer GaN.

\section{Conclusion}
In conclusion, using first-principle calculations, we predicted that both Ge and Sn atoms are likely to replace the Ga atoms of monolayer GaN, forming a buckling structure with a buckling height of 0.6 \AA\ and  0.9 \AA\, respectively. 
The diffusion barriers of the Ge atom along the direction vertical and parallel to the monolayer GaN are 0.12 eV and 0.82 eV, respectively. The corresponding diffusion barriers for the Sn atom are 0.54 eV and 0.62 eV, respectively.
Single Ge- and Sn-doped monolayer GaN exhibit as a semiconductor with a band gap of 0.40 and 0.58 eV, respectively. 
Both Ge and Sn dopants can exhibit a magnetic moment of 1 $\mu_B$ which prefers lying in the $xy-$plane.
The in-plane MAE is -14.55 $\rm \mu eV$ and -121.27 $\rm \mu eV$ for Ge and Sn dopant, respectively.
Compressive strain enhances the magnetic moment and MAE of the group-VI dopant, while a large tensile strain will destroy the magnetic moment of dopant.
Sn is better than Ge for the introduction of magnetic properties into monolayer GaN due to larger SOC, MAE, vertical diffusion barrier, and strain range to maintain the magnetic moment.  
As the doping concentration increases, the large vertical diffusion barrier to the monolayer GaN can be used to make both Ge and Sn atoms stay on the same side of monolayer GaN with an antiferromagnetic coupling between them.
Both N vacancies and Ga vacancies will remove the magnetic moment of Ge or Sn dopants. 
The plentiful supply of Ge or Sn dopants and the N-rich growth conditions help to realize the introduction of the magnetic properties into monolayer GaN for the spintronic applications. 

\section{Acknowledgments}
This work was supported by National Natural Science Foundation of China (No. 11904313 and 11904312),
the Scientific Research Foundation of the Higher Education of Hebei Province, China (No. BJ2020015),
the Natural Science Foundation of Hebei Province (No. A2020203027),
the Doctor Foundation Project of Yanshan University (No. BL19008).
Innovation Capability Improvement Project of Hebei province (No. 22567605H).
The numerical calculations in this paper have been done 
on the supercomputing system in the High Performance Computing Center 
of Yanshan University

\clearpage
\appendix
\section{}

\begin{figure}[htb!]
	\centerline{\includegraphics[width=0.9\textwidth]{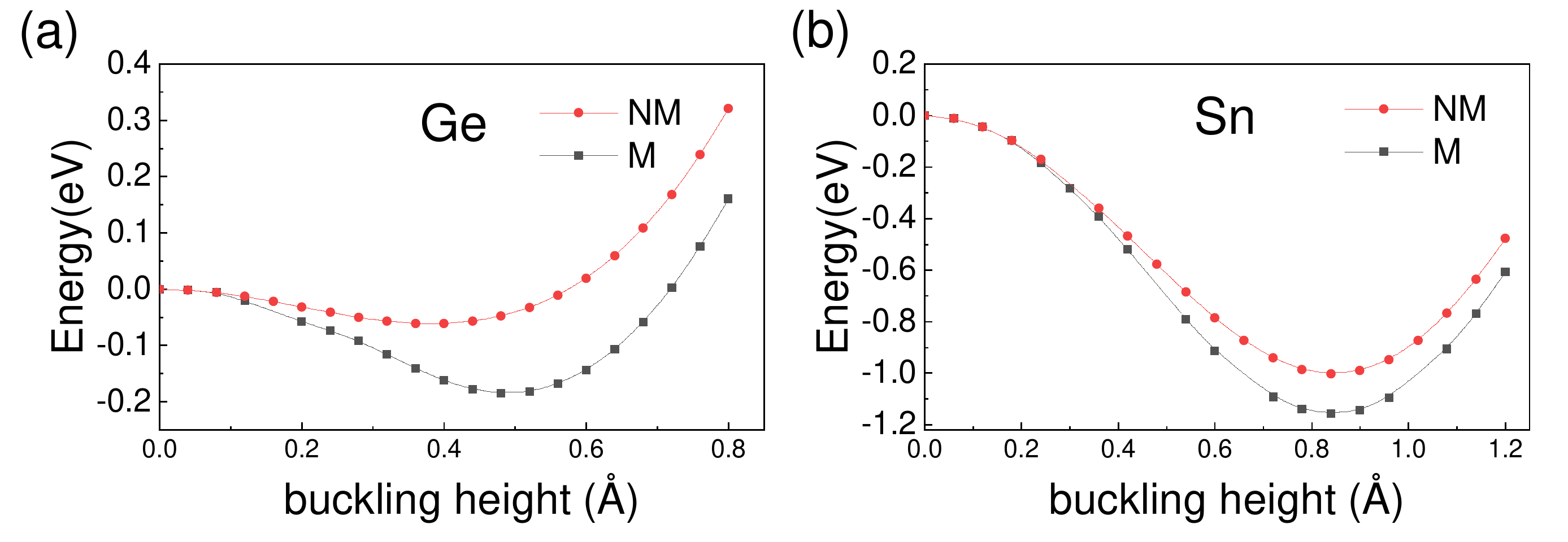}}
	\caption{The energy of (a) ${\rm Ge}_{\rm Ga,buckling}$ and (b) ${\rm Sn}_{\rm Ga,buckling}$ as a function of buckling height in non-magnetic (NM) and magmetic (M) state. The energy of planar doping structure was set to be zero. Here, the energies of buckling structure were calculated without structural relaxation of atomic position.}
	\label{a1} 
\end{figure}

\begin{figure}[htb!]
	\centerline{\includegraphics[width=0.9\textwidth]{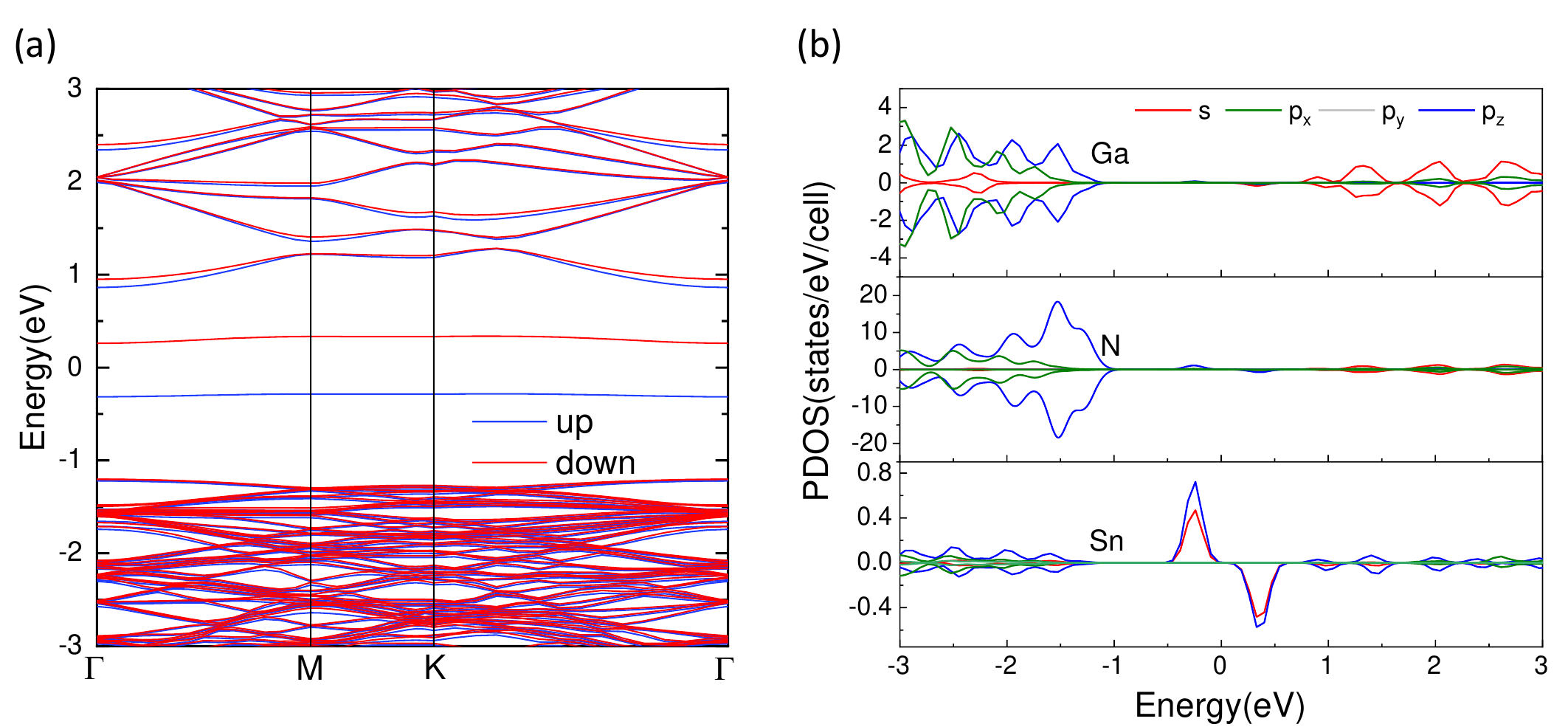}}
	\caption{(a) Energy bands and (b) PDOS of ${\rm Sn}_{\rm Ga,buckling}$ configuration. }
	\label{a2} 
\end{figure}

%\begin{figure}[htb!]
%	\centerline{\includegraphics[width=0.4\textwidth]{figureA3.pdf}}
%	\caption{(a) The orbital-resolved MAE of ${\rm Ge}_{\rm Ga,buckling}$ configuration. }
%	\label{a3} 
%\end{figure}

\begin{figure}[htb!]
	\centerline{\includegraphics[width=0.9\textwidth]{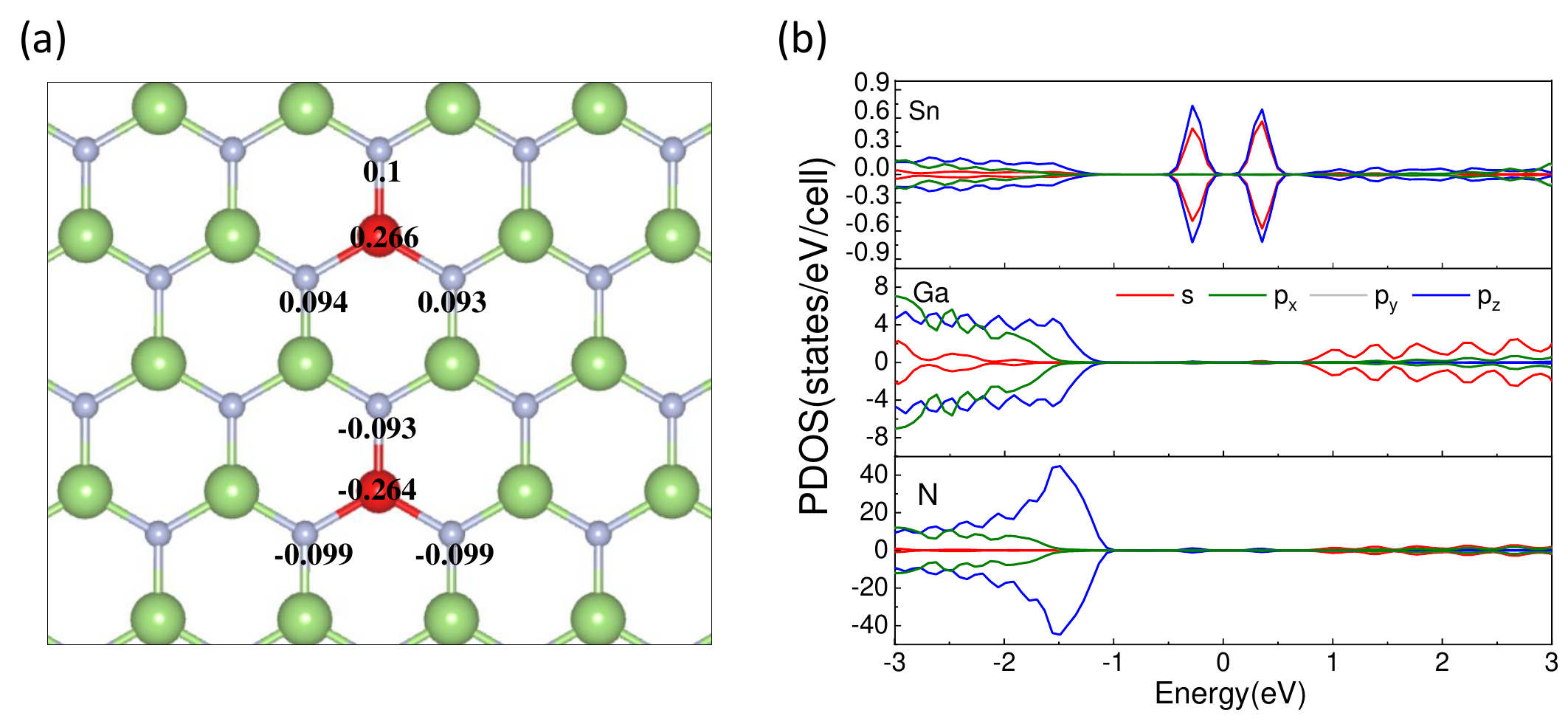}}
	\caption{(a)The main magnetic moment distribution of monolayer GaN doped with double Sn atoms in the same side of monolayer GaN. (b) The corresponding PDOS of configuration as shown in (a).}
	\label{a4}
\end{figure}

%
%\begin{center}
%{\footnotesize {\bf Table A1.} The energy generation (eV) of double atoms doping of $\rm V_{Ga}$\\
%\vspace{2mm}
%\begin{tabular}{ccc}
%\br
%{Position}                  & {Ga-rich} & {N-rich}  \\
%\mr
%{$\rm V_{Ga}-\rm Ge_{1}$} 	&-0.24	&-0.11\\
%{$\rm V_{Ga}-\rm Ge_{2}$} 	&0.04	&0.17\\
%{$\rm V_{Ga}-\rm Ge_{3}$}   &0.18	&0.32\\
%{$\rm V_{Ga}-\rm Ge_{4}$} 	&0.04	&0.17\\
%{$\rm V_{Ga}-\rm Ge_{5}$}   &0.18	&0.32\\
%{$\rm V_{Ga}-\rm Sn_{1}$} 	&-1.47	&-1.34\\
%{$\rm V_{Ga}-\rm Sn_{2}$} 	&-0.45 	&-0.31\\
%{$\rm V_{Ga}-\rm Sn_{3}$} 	&-0.23	&-0.10\\
%{$\rm V_{Ga}-\rm Sn_{4}$} 	&-0.45 	&-0.31\\
%{$\rm V_{Ga}-\rm Sn_{5}$} 	&-0.23	&-0.10\\
%\br
%\end{tabular}}
%\label{ta1}
%\end{center}

\newpage

\bibliographystyle{unsrt}
\bibliography{gan}

\begin{thebibliography}{10}

\bibitem{b8}
T.~Jungwirth, Jairo Sinova, J.~Ma\ifmmode~\check{s}\else \v{s}\fi{}ek,
  J.~Ku\ifmmode~\check{c}\else \v{c}\fi{}era, and A.~H. MacDonald.
\newblock {Theory of ferromagnetic (III,Mn)V semiconductors}.
\newblock {\em Rev. Mod. Phys.}, 78:809--864, Aug 2006.

\bibitem{b9}
T~Dietl and H~Ohno.
\newblock {Ferromagnetism in III–V and II–VI semiconductor structures}.
\newblock {\em Physica E}, 9(1):185--193, 2001.

\bibitem{Cui2005}
X.~Y. Cui, J.~E. Medvedeva, B.~Delley, A.~J. Freeman, N.~Newman, and
  C.~Stampfl.
\newblock Role of embedded clustering in dilute magnetic semiconductors: {Cr
  Doped GaN}.
\newblock {\em Phys. Rev. Lett.}, 95:256404, Dec 2005.

\bibitem{Zhang2021}
Chenwei Zhang, Penghui Yin, Wenhuan Lu, Victor Galievsky, and Pavle~V.
  Radovanovic.
\newblock On the origin of $d_0$ magnetism in transparent metal oxide
  nanocrystals.
\newblock {\em J. Phys. Chem. C}, 125(50):27714--27722, 2021.

\bibitem{Chakraborty2015}
Brahmananda Chakraborty and Lavanya~M. Ramaniah.
\newblock Exploring $d_0$ magnetism in doped {SnO$_2$}–a first principles
  {DFT} study.
\newblock {\em J. Magn. Magn. Mater.}, 385:207--216, 2015.

\bibitem{Coey2019}
J.~M.~D. Coey.
\newblock Magnetism in $d_0$ oxides.
\newblock {\em Nat. Mater.}, 18(7):652--656, 2019.

\bibitem{Huang2017}
Bevin Huang, Genevieve Clark, Efrén Navarro-Moratalla, Dahlia~R. Klein, Ran
  Cheng, Kyle~L. Seyler, Ding Zhong, Emma Schmidgall, Michael~A. McGuire,
  David~H. Cobden, Wang Yao, Di~Xiao, Pablo Jarillo-Herrero, and Xiaodong Xu.
\newblock Layer-dependent ferromagnetism in a van der waals crystal down to the
  monolayer limit.
\newblock {\em Nature}, 546(7657):270--273, 2017.

\bibitem{Gong2017}
Cheng Gong, Lin Li, Zhenglu Li, Huiwen Ji, Alex Stern, Yang Xia, Ting Cao, Wei
  Bao, Chenzhe Wang, Yuan Wang, Z.~Q. Qiu, R.~J. Cava, Steven~G. Louie, Jing
  Xia, and Xiang Zhang.
\newblock Discovery of intrinsic ferromagnetism in two-dimensional van der
  waals crystals.
\newblock {\em Nature}, 546(7657):265--269, 2017.

\bibitem{Ju2016}
Weiwei Ju, Tongwei Li, Zhiwei Hou, Hui Wang, Hongling Cui, and Xiaohong Li.
\newblock Exotic $d_0$ magnetism in partial hydrogenated silicene.
\newblock {\em Appl. Phys. Lett.}, 108(21):212403, 2016.

\bibitem{Gao2019}
Zijian Gao, Weiwei Ju, Tongwei Li, Qingxiao Zhou, Donghui Wang, Yi~Zhang, and
  Haisheng Li.
\newblock Tunable magnetism in defective {MoS$_2$} monolayer with nonmetal
  atoms adsorption.
\newblock {\em Superlattices Microstruct.}, 130:346--353, 2019.

\bibitem{Bai2015}
Yujie Bai, Kaiming Deng, and Erjun Kan.
\newblock Electronic and magnetic properties of an {AlN} monolayer doped with
  first-row elements: a first-principles study.
\newblock {\em RSC Adv.}, 5:18352--18358, 2015.

\bibitem{Han2017}
Ruilin Han, Xiaoyang Chen, and Yu~Yan.
\newblock Magnetic properties of {AlN} monolayer doped with group {1A or 2A}
  nonmagnetic element: First-principles study.
\newblock {\em Chin. Phys. B}, 26(9):097503, aug 2017.

\bibitem{a1}
Wenhui Wan, Na~Kang, Yanfeng Ge, and Yong Liu.
\newblock The theoretical study of unexpected magnetism in {2D Si-Doped AlN}.
\newblock {\em Front. Phys.}, 10:843352, 2022.

\bibitem{Xiao2018}
Wen-Zhi Xiao, Gang Xiao, Qing-Yan Rong, and Ling-Ling Wang.
\newblock Electronic and magnetic properties of {SnS$_2$} monolayer doped with
  non-magnetic elements.
\newblock {\em Physica E}, 99:182--188, 2018.

\bibitem{Adhikary2007}
Kalyan Adhikary and Subhadra Chaudhuri.
\newblock Gallium nitride: Synthesis and characterization.
\newblock {\em Trans. Indian Ceram. Soc.}, 66(1):1--16, 2007.

\bibitem{Jabbar2021}
Haneen~D. Jabbar, Makram~A. Fakhri, and Mohammed {Jalal AbdulRazzaq}.
\newblock Gallium nitride –based photodiode: A review.
\newblock {\em Mater. Today:. Proc.}, 42:2829--2834, 2021.

\bibitem{Paula2017}
Wesley~J. de~Paula, Pedro~L. Tavares, Denis De~C.~Pereira, Gabriel~M. Tavares,
  Filipe~L. Silva, Pedro~S. Almeida, and Henrique A.~C. Braga.
\newblock A review on gallium nitride switching power devices and applications.
\newblock In {\em 2017 Brazilian Power Electronics Conference (COBEP)}, pages
  1--7, 2017.

\bibitem{a6}
Xing~Xiang Ruan, Cansheng Huang, Fuchun Zhang, Hui Fang, and Weihu Zhang.
\newblock {First-principles study on electromagnetic properties of Mn-doped
  GaN}.
\newblock {\em Ferroelectrics}, 547:104 -- 97, 2019.

\bibitem{a8}
P.R. Hageman, W.J. Schaff, Jacek Janinski, and Zuzanna Liliental-Weber.
\newblock {n-type doping of wurtzite GaN with germanium grown with
  plasma-assisted molecular beam epitaxy}.
\newblock {\em J. Cryst. Growth}, 267(1):123--128, 2004.

\bibitem{a4}
S.~Fritze, A.~Dadgar, H.~Witte, M.~Bügler, A.~Rohrbeck, J.~Bläsing,
  A.~Hoffmann, and A.~Krost.
\newblock {High Si and Ge n-type doping of {GaN} doping - Limits and impact on
  stress}.
\newblock {\em Appl. Phys. Lett.}, 100(12):122104, 2012.

\bibitem{AlBalushi2016}
Zakaria~Y. Al~Balushi, Ke~Wang, Ram~Krishna Ghosh, Rafael~A. Vilá, Sarah~M.
  Eichfeld, Joshua~D. Caldwell, Xiaoye Qin, Yu-Chuan Lin, Paul~A. DeSario, Greg
  Stone, Shruti Subramanian, Dennis~F. Paul, Robert~M. Wallace, Suman Datta,
  Joan M. Redwing, and Joshua~A. Robinson.
\newblock Two-dimensional gallium nitride realized via graphene encapsulation.
\newblock {\em Nat. Mater.}, 15(11):1166--1171, 2016.

\bibitem{Koratkar2016}
Nikhil~A. Koratkar.
\newblock Two-dimensional gallium nitride.
\newblock {\em Nat. Mater.}, 15(11):1153--1154, 2016.

\bibitem{Li2018}
Jiabin Li and Hongxia Liu.
\newblock {Magnetism investigation of GaN monolayer doped with group VIII B
  transition metals}.
\newblock {\em J. Mater. Sci.}, 53(23):15986--15994, 2018.

\bibitem{a13}
Gang Xiao, Ling-Ling Wang, Qing-Yan Rong, Hai-Qing Xu, and Wen-Zhi Xiao.
\newblock {A comparative study on magnetic properties of Mo doped AlN, GaN and
  InN monolayers from first-principles}.
\newblock {\em Phys. B}, 524:47--52, 2017.

\bibitem{Hussain2015}
Fayyaz Hussain, Y.~Q. Cai, M.~Junaid~Iqbal Khan, Muhammad Imran, Muhammad
  Rashid, Hafeez Ullah, Ejaz Ahmad, Farhana Kousar, and S.~A. Ahmad.
\newblock Enhanced ferromagnetic properties of {Cu} doped two-dimensional {GaN}
  monolayer.
\newblock {\em Int. J. Mod. Phys. C}, 26(01):1550009, 2015.

\bibitem{a3}
Basanta Roul, Mohana~K. Rajpalke, Thirumaleshwara~N. Bhat, Mahesh Kumar, A.~T.
  Kalghatgi, S.~B. Krupanidhi, Nitesh Kumar, and A.~Sundaresan.
\newblock {Experimental evidence of Ga-vacancy induced room temperature
  ferromagnetic behavior in GaN films}.
\newblock {\em Appl. Phys. Lett.}, 99(16):162512, 2011.

\bibitem{a10}
Qian Zhao, Zhihua Xiong, Zhenzhen Qin, Lanli Chen, Ning Wu, and Xingxing Li.
\newblock {Tuning magnetism of monolayer GaN by vacancy and nonmagnetic
  chemical doping}.
\newblock {\em J. Phys. Chem. Solids}, 91:1--6, 2016.

\bibitem{a12}
Vo~{Van On}, J.~Guerrero-Sanchez, R.~Ponce-Pérez, and D.M. Hoat.
\newblock Study of vacancy, voids, atom adsorption and domain substitution in
  hexagonal gallium nitride monolayer.
\newblock {\em Surf. Interfaces}, 30:101898, 2022.

\bibitem{Ganchenkova2006}
M.~G. Ganchenkova and R.~M. Nieminen.
\newblock Nitrogen vacancies as major point defects in gallium nitride.
\newblock {\em Phys. Rev. Lett.}, 96:196402, May 2006.

\bibitem{a11}
Roberto González, William López-Pérez, Álvaro González-García, María~G.
  Moreno-Armenta, and Rafael González-Hernández.
\newblock {Vacancy charged defects in two-dimensional GaN}.
\newblock {\em Appl. Surf. Sci.}, 433:1049--1055, 2018.

\bibitem{Gao2017}
Han Gao, Han Ye, Zhongyuan Yu, Yunzhen Zhang, Yumin Liu, and Yinfeng Li.
\newblock Point defects and composition in hexagonal {group-III} nitride
  monolayers: A first-principles calculation.
\newblock {\em Superlattices Microstruct.}, 112:136--142, 2017.

\bibitem{Mu2015}
Yuewen Mu.
\newblock Chemical functionalization of {GaN} monolayer by adatom adsorption.
\newblock {\em J. Phys. Chem. C}, 119(36):20911--20916, 2015.

\bibitem{Tang2018}
Wencheng Tang, Minglei Sun, Jin Yu, and Jyh-Pin Chou.
\newblock Magnetism in non-metal atoms adsorbed graphene-like gallium nitride
  monolayers.
\newblock {\em Appl. Surf. Sci.}, 427:609--612, 2018.

\bibitem{Kadioglu2018}
Yelda Kadioglu, Fatih Ersan, Deniz Kecik, Olcay~Üzengi Aktürk, Ethem Aktürk,
  and Salim Ciraci.
\newblock Chemical and substitutional doping{,} and anti-site and vacancy
  formation in monolayer {AlN and GaN}.
\newblock {\em Phys. Chem. Chem. Phys.}, 20:16077--16091, 2018.

\bibitem{a9}
Naresh Alaal and Iman~S. Roqan.
\newblock Tuning the electronic properties of hexagonal two-dimensional {GaN}
  monolayers via doping for enhanced optoelectronic applications.
\newblock {\em ACS Appl. Nano Mater.}, 2(1):202--213, 2019.

\bibitem{Yadav2022a}
Sandeep Yadav, B.K. Agrawal, and P.S. Yadav.
\newblock Non-magnetic adsorbent functionalized magnetism and spin filtering in
  a two-dimensional {GaN} monolayer.
\newblock {\em J. Phys. Chem. Solids}, 167:110731, 2022.

\bibitem{Gupta2014}
Sanjeev~K. Gupta, Haiying He, Douglas Banyai, Mingsu Si, Ravindra Pandey, and
  Shashi~P. Karna.
\newblock {Effect of Si doping on the electronic properties of BN monolayer}.
\newblock {\em Nanoscale}, 6:5526--5531, 2014.

\bibitem{b2}
G.~Kresse and J.~Furthm\"uller.
\newblock {Efficient iterative schemes for ab initio total-energy calculations
  using a plane-wave basis set}.
\newblock {\em Phys. Rev. B}, 54:11169--11186, Oct 1996.

\bibitem{b3}
G.~Kresse and D.~Joubert.
\newblock {From ultrasoft pseudopotentials to the projector augmented-wave
  method}.
\newblock {\em Phys. Rev. B}, 59:1758--1775, Jan 1999.

\bibitem{b4}
John~P Perdew, Kieron Burke, and Matthias Ernzerhof.
\newblock Generalized gradient approximation made simple.
\newblock {\em Phys. Rev. Lett.}, 77(18):3865, 1996.

\bibitem{b5}
Hendrik~J Monkhorst and James~D Pack.
\newblock {Special points for Brillouin-zone integrations}.
\newblock {\em Phys. Rev. B}, 13(12):5188, 1976.

\bibitem{a14}
X.~Y. Cui, B.~Delley, and C.~Stampfl.
\newblock {Band gap engineering of wurtzite and zinc-blende GaN/AlN
  superlattices from first principles}.
\newblock {\em J. Appl. Phys.}, 108(10):103701, 2010.

\bibitem{a15}
Heinz Schulz and K.H. Thiemann.
\newblock {Crystal structure refinement of AlN and GaN}.
\newblock {\em Solid State Commun.}, 23(11):815--819, 1977.

\bibitem{Mills1995}
Gregory Mills, Hannes Jónsson, and Gregory~K. Schenter.
\newblock Reversible work transition state theory: application to dissociative
  adsorption of hydrogen.
\newblock {\em Surf. Sci.}, 324(2):305--337, 1995.

\bibitem{Fan2022}
Xiaotian Fang, Baozeng Zhou, Xiaocha Wang, and Wenbo Mi.
\newblock High curie temperature and large perpendicular magnetic anisotropy in
  two-dimensional half metallic osi3 monolayer with quantum anomalous hall
  effect.
\newblock {\em Mater. Today Phys.}, 28:100847, 2022.

\bibitem{wang2023}
Yuwan Wang, Xicheng Zhang, Zhihao Gao, Tengfei Cao, Junqin Shi, and Xiaoli Fan.
\newblock Ferromagnetic gdx (x = cl, br) monolayers with large perpendicular
  magnetic anisotropy and high curie temperature.
\newblock {\em J. Phys. Chem. C}, 127(9):4643--4650, 2023.

\bibitem{shengmei2021}
Shengmei Qi, Jiawei Jiang, Xiaocha Wang, and Wenbo Mi.
\newblock Valley polarization, magnetic anisotropy and dzyaloshinskii-moriya
  interaction of two-dimensional graphene/janus 2h-vsex (x = s, te)
  heterostructures.
\newblock {\em Carbon}, 174:540--555, 2021.

\bibitem{yan2021}
Shiming Yan, Wen Qiao, Deyou Jin, Xiaoyong Xu, Wenbo Mi, and Dunhui Wang.
\newblock Role of exchange splitting and ligand-field splitting in tuning the
  magnetic anisotropy of an individual iridium atom on
  $\mathrm{Ta}{\mathrm{s}}_{2}$ substrate.
\newblock {\em Phys. Rev. B}, 103:224432, Jun 2021.

\bibitem{zhang2020}
Fang Zhang, Wenbo Mi, and Xiaocha Wang.
\newblock Spin-dependent electronic structure and magnetic anisotropy of 2d
  ferromagnetic janus cr2i3x3 (x = br, cl) monolayers.
\newblock {\em Adv. Electron. Mater.}, 6(1):1900778, 2020.

\bibitem{PRB1993}
Ding-sheng Wang, Ruqian Wu, and A.~J. Freeman.
\newblock First-principles theory of surface magnetocrystalline anisotropy and
  the diatomic-pair model.
\newblock {\em Phys. Rev. B}, 47:14932--14947, Jun 1993.

\bibitem{Gerrit1998}
Gerrit van~der Laan.
\newblock Microscopic origin of magnetocrystalline anisotropy in transition
  metal thin films.
\newblock {\em J. Phys.: Condens. Matter}, 10(14):3239, apr 1998.

\bibitem{Tkachenko2020}
Nikolay~V. Tkachenko, Bingyi Song, Dmitriy Steglenko, Ruslan~M. Minyaev,
  Li-Ming Yang, and Alexander~I. Boldyrev.
\newblock Computational prediction of the low-temperature ferromagnetic
  semiconducting {2D SiN} monolayer.
\newblock {\em Phys Status Solidi B}, 257(3):1900619, 2020.

\bibitem{Cui2019}
Hao Cui, Dachang Chen, Chao Yan, Ying Zhang, and Xiaoxing Zhang.
\newblock {Repairing the N-vacancy in an InN monolayer using NO molecules: a
  first-principles study}.
\newblock {\em Nanoscale Adv.}, 1:2003--2008, 2019.

\end{thebibliography}

\end{document}